\documentclass[prb,twocolumn]{revtex4}
\usepackage[latin1]{inputenc}
\usepackage{latexsym,amstext,amssymb,amsmath}
\usepackage[dvips]{graphicx}
\usepackage{color}
\begin{document}
%%%%%%%%%%%%%%%%%%%%%%%%%%%%%%%%%%%%%%%%%%%%%%%%%%%%%%%%%%%%%%%%%%%%%%%%%%%

\title{Role of shuffles and atomic disorder in Ni-Mn-Ga}

\author{A. T. Zayak}
\email{alexei@thp.uni-duisburg.de}
\author{P. Entel}
\affiliation{Institute of Physics, University Duisburg-Essen,
             47048 Duisburg, Germany}

\date{\today}

\begin{abstract}

We report results of \textit{ab-initio} calculations of the ferromagnetic
Heusler alloy  Ni-Mn-Ga. 
Particular emphasis is placed on the stability of the low temperature
tetragonal structure with $c/a = 0.94$. 
This structure cannot be derived from the parent L2$_1$ structure
by a simple homogeneous strain associated with the soft elastic constant $C'$. 
In order to stabilise the tetragonal phase, one has to take into account
shuffles of atoms, which form a wave-like pattern of atomic displacements
with a well defined period (modulation).    
While the modulation is related to the soft acoustic [110]-TA$_2$ phonon
mode observed in Ni$_2$MnGa, we obtain
additional atomic shuffles, which are related to acoustic-optical
coupling of the phonons in Ni$_2$MnGa. In addition, we have simulated an
off-stoichiometric systems, in which 25 \% of Mn atoms are replaced by Ni. 
The energy of this structure also exhibits a local minimum at
$c/a = 0.94$. This allows us to conclude that both shuffles and atomic
disorder stabilize the $c/a = 0.94$ structure. In both cases
the stability seems to be associated with a dip in the minority-spin density of
states (DOS) at the Fermi level, being related to the formation
of hybrid states of Ni-\textit{d} and Ga-\textit{p} minority-spin orbitals.\\

\textbf{Keywords:} Ni-Mn-Ga alloys, martensitic phase, modulated structure

\vspace{5mm}
\begin{center}
 (Accepted for Materials Science and Engineering A)
\end{center}

\end{abstract}

%\textbf{Keywords:} Ni-Mn-Ga alloys, martensitic phase, modulated structure

\vspace{0.5cm}

\maketitle

Ni-Mn-Ga alloys (close to stoichiometric Ni$_2$MnGa) are known to exhibit
unique magneto-elastic properties. They are ferromagnetic at room
temperature ($T_{\mathrm{C}} \sim 380 \; K$), and undergo
(including the precursor) a two-step martensitic  transformation
for $T_M < T_C$ ($T_{\mathrm{M}} \sim 200 \; K$) 
[1]. In the
martensitic state of Ni-Mn-Ga the structure consists of differently oriented
martensitic domains (twin variants), which are also magnetic domains.
This makes the martensitic structure of Ni-Mn-Ga sensible to an external
magnetic field, which can induce a redistribution of the martensitic domains
in the sample. Those domains with easy magnetic axes are along the field will
gain in energy on cost of the domains with less favourable orientation of the
magnetisation. Alignment of twin variants by the motion of twin
boundaries can result in large macroscopic strain up to 6 \% [2, 3]. 

This effect is used in the magnetic-shape-memory (MSM) technology
[4]. The MSM technology is based on the magnetic field induced
redistribution of martensitic domains in the sample. From a technological
point of view, Ni-Mn-Ga is more promising than other materials being
presently in commercial use,  for example, the well-known material Tb-Dy-Fe
(Terfenol-D) which exhibits magnetostrictive strains of about 0.1 \%.
Design of new efficient MSM magneto-mechanical actuator devices is in progress
[5].

In this work we discuss the stability of different
structures inside a single martensitic variant of Ni-Mn-Ga. From experimental 
studies we know that these alloys can form at least three different phases
in the martensitic state [6]. Depending mostly on composition, crystals
can be found in the modulated tetragonal structure with $c/a = 0.94$ known as
5M or 10M, the orthorhombic modulated structure 7M, or the tetragonal structure
with $c/a \approx 1.2$. Sometimes, the modulated structures are also denoted
by 5R and 7R [7].

The central question concerns the nature of the modulation in the 5M and
7M structures, which has often not often taken into account in theoretical
investigations [8-10]. However, the modulation plays a basic role in
Ni$_2$MnGa, 
as shown by recent first-principles calculations [11]. In the literature, it
is argued that the reason for the modulation arises
from specific nesting properties of the Fermi surface in Ni$_2$MnGa, which in 
turn, causes softening of the acoustic [110]-TA$_2$ phonon mode [12].  

In the previous study [11] of the stability of the 5M structure, 
we have drawn attention to specific changes in the DOS, namely a dip,
which develops in the minority-spin density of states (DOS) at the Fermi
level, being related to the Fermi surface geometry [13].
Analysis of the partial contributions to the DOS shows that this dip is formed
by Ni-\textit{d}$\downarrow$ and Ga-\textit{p}$\downarrow$ electrons. Hence,
there is a strong evidence that the formation of alike hybrid states might
contribute to the stability of the modulated structures. In addition to the
shuffles leading to the modulation, atomic disorder and non-collinearity of
magnetic moments were proposed as possible factors being also responsible for
the structural stability of the modulated phases in Ni-Mn-Ga.  

In this work, we discuss properties of the modulated 5M
structure with new features which have been obtained in recent
calculations. In addition, we consider an off-stoichiometric composition of
Ni-Mn-Ga with excess Ni replacing Mn atoms. 
For the total energy calculation we use the 
Vienna {\it Ab-initio} Simulation Package (VASP) [14,15] and the
implemented  projector-augmented wave formalism (PAW) [16].
The electronic exchange and correlations are treated by the generalised
gradient approximation (GGA). The 3{\it d} electrons of Ga have
been included as valence states.

\begin{figure}[t]
  \centering
  \resizebox{8cm}{!}{\includegraphics*{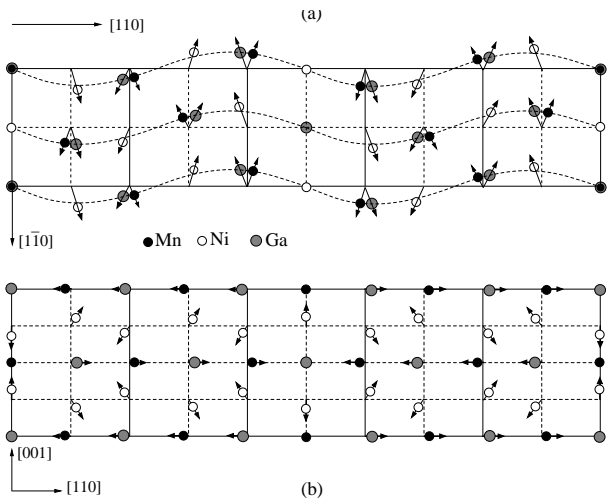}}
\caption{\label{planes} Modulated martensitic structure of Ni$_2$MnGa (5M)
  shown schematically: (a) projection of the 5M structure on the (001) plane
  (top view) and (b) projection of the 5M structure on the (1\=10) plane (side
  view). Filled, open and gray dots show the positions of Mn, Ni and Ga,
  respectively. 
  In addition to the modulation, which moves the atoms along the [1\=10]
  direction only, there is a tiny shuffling of the atoms of the order of 0.001
  \AA\ ($\sim$ 1 \% 
  of the modulation amplitude), which moves the atoms in [001] and [110]
  directions. The extremely small displacements of atoms
  have been enlarged in the figure for clearness.}
\end{figure}

In order to simulate the 5M structure, we used a supercell, which resembles
the five-layered structure obtained by V. V. Martynov and V. V. Kokorin in
experiment [6]. The supercell consists of five tetragonal unit cells of
Ni$_2$MnGa, similar to those used in Ref. [9]. This allows us to incorporate
the full period of the 5M 
modulation in the supercell [11]. The basal plane is spanned by [1\={1}0] and
[001] of the $\mathrm L2_{1}$ structure with the modulation along [110].
Altogether we use ten atomic planes perpendicular to [110] in order to form the
supercell. The modulation is generated by displacing these atomic planes along
[1\={1}0] direction. By this construction, two full five-layered periods
fit into the supercell. The initial magnitudes of the atomic displacements
were chosen according to the 5M structure of Ref. [11], for the Ni-plane
equal to 0.324 \AA\ and for the Ga-Mn-plane equal to 0.292 \AA.

This  supercell has an orthorhombic symmetry with lattice parameters 
$a \approx 4.17$ \AA, $b \approx 20.73$ \AA, $c \approx
5.633$ \AA, yielding a tetragonality ratio of $c/a
\approx 0.955$, i.e, we take exactly the structure which was obtained
theoretically in Ref. [11]. In the calculations we allowed for relaxation of
the structure by changes of the volume, cell 
shape and atomic positions. The plane-wave cutoff energy was
equal to 241.6 eV and the k-points were generated using the Monkhorst-Pack
method with a grid of \ 10 $\times$ 2 $\times$ 8\ points in the full
Brillouin zone for the long orthorhombic supercell.
Actually, all parameters of the calculations are the same as
used in our previous calculations [11], except for the k-points mesh.

\begin{figure}[t]
  \centering
  \resizebox{8cm}{!}{\includegraphics*{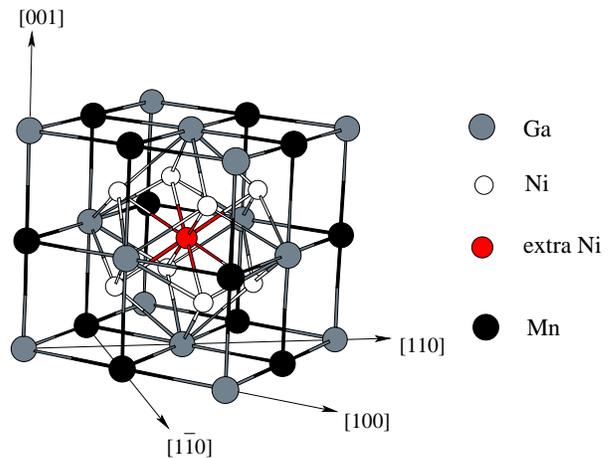}}
\caption{\label{struc} Off-stoichiometric supercell of Ni$_{2+x}$Mn$_{1-x}$Ga
  for $x=0.25$ created by replacing Mn by an Ni atom in one of the 16 positions
of the L2$_1$ structure.}
\end{figure}

The structure obtained after relaxation is shown in Fig. 1a and b. 
As in the model for the 5M structure in Ref. [6,11], 
the modulation has form of a static wave
with polarization (the direction in which the atoms move) along [1\=10], which
is perpendicular to the propagation direction [110]. All atoms move with the
same phase that agrees with the acoustic mode [12,17].
However, a careful analysis shows that the shuffling of the atoms in
Ni$_2$MnGa consists of two different contributions, being a superposition of
them. The first one is the \textit{modulation} which is wave-like.
While the second one is different in the sense that it is not a wave. We call
these additional shuffles \textit{tetrahedral distortions}, similar to those
considered by Harrison for tetrahedrally coordinated solids having the
symmetry of diamond or zincblende [18].
According to Harrison, radial and angular distortions of tetrahedral
structures stand for this motion of the atoms. In case of Ni$_2$MnGa each Ni
atom is surrounded by two tetrahedrals formed by Ga and Mn, respectively. Note
that amplitudes of the tetrahedral distortions are of the order of 
0.001 \AA\ which is $\sim$1\% of the modulation amplitude. 

We would like to draw attention to some features related to the tetrahedral
distortions. 
In the (001) plane, Fig. 1a, the Ga and Ni atoms move closer to each other,
while the Mn atoms move towards the free space left by the Ni atoms. Also, the
plane 
(1\=10) in Fig. 1(b) shows couples of Ni atoms moving closer to their nearest
Ga atom, while the nearest to them Mn atom is pushed out.
Thus, each atom of Ga tends to couple with two
Ni atoms, which will contribute to the
hybridization of the Ga-\textit{p} and Ni-\textit{d} orbitals as discussed
above. We stress that the modulation and the tetrahedral distortions 
must be considered separately from each other. The first one is related to the
soft acoustic 
[110]-TA$_2$ phonon mode, while the second one is connected with the
coupling of acoustic phonon modes [110]-TA$_1$ and LA with
optical modes derived mostly from the Ni atoms, as it was discussed in
Ref. [17].

\begin{figure}[t]
  \centering
  \resizebox{8cm}{!}{\includegraphics*{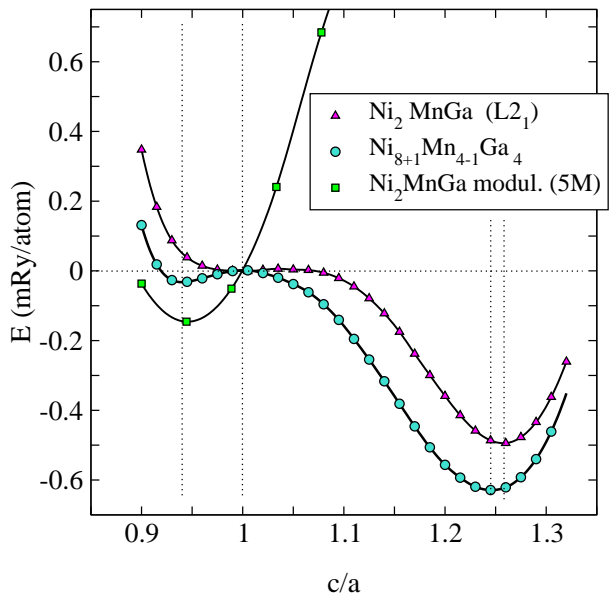}}
\caption{\label{ca} The relative change of the total energy of Ni-Mn-Ga as a
  function of $c/a$ for the three different cases: cubic Ni$_2$MnGa structure,
modulated 5M structure and off-stoichiometric case defined by
 Ni$_{8+1}$Mn$_{4-1}$Ga$_4$.}
\end{figure}

As a matter of fact, most experimental studies concentrate on
off-stoichiometric compositions of Ni-Mn-Ga, while theoretical studies are
mostly done for for the stoichiometric structure Ni$_2$MnGa. Hence,
theoretical investigations for off-stoichiometric compositions are urgently
needed. The results obtained so far allow to discuss importance of the   
 Ni-Ga interaction in Ni-Mn-Ga, showing that the stability of the modulated
 shuffles benefits from this coupling (and the locally broken symmetry). The
modulation facilitates the formation of hybrid Ga-\textit{p} states 
with other atoms, which is less favourable in a perfectly symmetric
environment, 
where Ga ``cannot choose'' neighbours to couple (the shuffles of atoms remove
this degeneracy allowing for a lower energy of the system).

The question now is whether the modulated shuffles are the only way to remove
the perfect order. This could also be done by doping the structure by
local defects. Planing to check this we have simulated off-stoichiometric
 Ni-Mn-Ga structures.  
In order to simulate an off-stoichiometric composition,
a cubic supercell of 16 atoms is used in the calculations. The corresponding
structure with a composition formula Ni$_{8+1}$Mn$_{4-1}$Ga$_4$ is shown in
Fig. 2. An extra atom of Ni is placed into a Mn position  
(central atom in Fig. 2). The supercell has been optimized, first, with
respect to the atomic positions and the volume. The $c/a = 1$ ratio was kept
constant at this stage of the calculations. After the relaxation, the optimal
volume became smaller (192.51 \AA$^3$) compared to the stoichiometric
structure (195.79 \AA$^3$) by about 1.6\%. The atomic positions have changed
as well, but only for the Ni atoms. The positions of Mn, Ga and the extra-Ni
remained unchanged, which is not physical, but is due to the choice of the
supercell and the periodic boundary conditions used for the calculations. 
One would need to simulate a bigger supercell for a more sophisticatd
treatment of the defects in Ni-Mn-Ga. In
Fig. 2 the extra-Ni is shown in the center connected by bonds to 
its eight nearest Ni neighbours. The relaxation has led to
considerable displacements of all eight Ni atoms towards the extra-Ni
along the bonds, while the rest of the structure remains
unchanged. The magnetic moments remained ferromagnetically aligned. Although,
the 
extra-Ni  has got a smaller magnetic moment ($0.24 \; \mu_{B}$) compared to
the regular Ni atoms ($\approx 0.35  \; \mu_{B}$).

This relaxed cubic structure is now used to study the impact of a
tetragonal deformation. Fig. 3 shows the 
dependence of the total energy on the $c/a$ ratio for the three cases: the
perfect stoichiometric Ni$_2$MnGa structure, the 5M structure, and the
off-stoichiometric structure defined by Ni$_{8+1}$Mn$_{4-1}$Ga$_4$. 
It turns out that the total energy curve for the case of the
off-stoichiometry  has a minimum at $c/a = 0.94$, exactly the  value 
as for the 5M structure, and in the 
experimental investigations. This shows that both modulation and disorder
are responsible for the stable tetragonal structure with $c/a = 0.94$.

Fig. 4 presents the total electronic density of states for the
off-stoichiometric case (structure shown in Fig. 2). The DOS is shown for: the
cubic structure with $c/a = 1$ and the tetragonal structure with 
$c/a = 0.94$. Important is here the dip in the
minority-spin density of states right at the Fermi level. The dip is present
for 
both cases, but stronger developed for the more stable tetragonal
structure. The analysis of the partial DOS shows again that the two peaks,
responsible for the dip, originate from the Ni-\textit{d} and
Ga-\textit{p} orbitals (the effect is similar to the results obtained
for the 5M structure in Ref. [11]).

\begin{figure}[t]
  \centering
  \resizebox{8cm}{!}{\includegraphics*{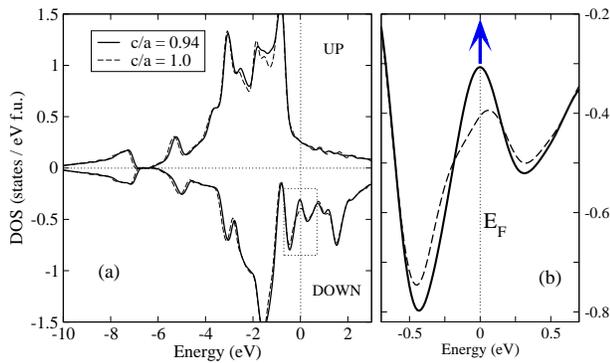}}
\caption{\label{dos} Electronic density of states of the off-stoichiometric
  alloy Ni$_{2+x}$Mn$_{1-x}$Ga with $x=0.25$  for the two cases: unstable $c/a
  = 1$ and metastable $c/a = 0.94$. (a) Total DOS for the cell with 16
  atoms. (b) Enlargement of the total DOS which shows closer the dip in the
  minority spin density of states. Analysis of the partial density of states
  shows that the states which form the dip are derived from the Ni-\textit{d}
  and Ga-\textit{p} orbitals.}
\end{figure}

In summary, we have presented computational results for the modulated 5M
structure of Ni$_2$MnGa. We find that shuffles in Ni$_2$MnGa involve two
different effects, which contribute to the pattern of the atomic
displacements. The first and the stronger contribution, called modulation,
arises from the soft acoustic [110]-TA$_2$ phonon mode, which is observed in
the phonon spectrum of Ni$_2$MnGa. The second effect, referred in our work as
tetrahedral distortions, yields small (as compared to the modulation)
atomic displacements and results from the coupling between the acoustic
[110]-TA$_1$ and LA with corresponding low-energy optical modes of Ni. The
calculations for the off-stoichiometric Ni$_{8+1}$Mn$_{4-1}$Ga$_4$ supercell
yield a local minimum of the total energy at $c/a = 0.94$, which is the
``natural'' tetragonality ratio for the Ni-Mn-Ga at low temperatures. The same
tetragonality ratio can be obtained in the calculations when taking 
modulation (5M structure) into account. Similar to the case of the 5M
structure, the stability of the $c/a = 0.94$ ratio is related to a dip in the
minority-spin electron density of states which develops from 
hybridizing Ni-\textit{d} and 
Ga-\textit{p} states right at the Fermi level, stabilizing the
tetragonal variant with $c/a = 0.94$. In other words, the tetragonal structure
becomes stable due to the covalent interaction of Ni an Ga atoms via the p-d
hybrid electronic orbitals, the importance of which for the Heusler alloys
have been duscussed by K\"ubler \textit{et al.} [19]. But we emphasize the
role of the local symmetry loss which facilitates the formation of the p-d
hybrid states. 
In future work we will perform calculations using the
GGA+U method. This will split apart the
bonding and anti-bonding states around the Fermi level, which will
further deepen the dip discussed above. 
We expect that the resulting total energy will correspond then to the 5M 
structure as the ground state.

\section*{Acknowledgements}
This work has been supported by the Graduate School ``Structure and
Dynamics of Heterogeneous Systems'' of the Deutsche Forschungsgemeinschaft
(DFG). We thank Dr. A. Postnikov, Dr. A. Ayuela and Dr. J. Enkovaara for
valuable discussions.

\vspace{1cm}

\section*{References}

\noindent
[1]
P. J. Webster, K. R. A. Ziebeck, S. L. Town, and M. S. Peak,
Phil. Mag. \textbf{49} (1984) 295

\noindent
[2]
R.C. O'Handley,
J. Appl. Phys. \textbf{83} (1998) 3263

\noindent
[3]
R. C. O`Handley, S. J. Murray, M. Marioni,  
            H. Nembach, and S. M. Allen,
J. Appl. Phys. \textbf{87} (2000) 4712

\noindent
[4]
K. Ullakko, J. K. Huang, C. Kantner, R. C. O'Handley, and
V. V. Kokorin,
 Appl. Phys. Lett. \textbf{69} (1996) 1966

\noindent
[5]
I. Aaltio and K. Ullakko,
Proceedings of the 7th International Conferenc on 
       New Actuators, {ACTUATOR 2000}, Bremen Germany, June 2000, p.45

\noindent
[6]
V. V. Martynov and V. V. Kokorin,
J. Phys. III (France) \textbf{2} (1992) 739

\noindent
[7]
Lluis Manosa, Antoni Planes, J. Zarestky,
       T. Lograsso, D.L. Schlagel and C. Stassis,
Phys. Rev. B \textbf{64} (2001) 024305

\noindent
[8]
V. V. Godlevsky and K. M. Rabe,
Phys. Rev. B \textbf{63} (2001) 134407

\noindent
[9]
A. Ayuela, J. Enkovaara, K. Ullakko, and R. M. Nieminen,
J. Phys.: Condens. Matter \textbf{11} (1999) 2017

\noindent
[10]
A. T. Zayak, P. Entel, and J. Hafner,
J. Phys.IV (France): International conference on martensitic
       transformations ICOMAT'02, 2003, to appear

\noindent
[11]
A. T. Zayak, P. Entel, J. Enkovaara, A. Ayuela, and R. M. Nieminen,
J. Phys.: Condens. Matter \textbf{15} (2003) 159

\noindent
[12]
A. Zheludev, S. M. Shapiro, P. Wochner,
       and L. E. Tanner,
Phys. Rev. B \textbf{54} (1996) 15045

\noindent
[13]
Yongbin Lee, Joo Yull Rhee, and B. N. Harmon,
Phys. Rev. B \textbf{66} (2002) 054424

\noindent
[14]
G. Kresse and J. Furthm\"uller,
Phys. Rev. B \textbf{54} (1996) 11169

\noindent
[15]
G. Kresse and D. Joubert,
Phys. Rev. B \textbf{59} (1999) 1758

\noindent
[16]
Peter Blöchl,
Phys. Rev. B \textbf{50} (1994) 17953

\noindent
[17]
A. T. Zayak, P. Entel, J. Enkovaara, A. Ayuela,
      and R. M. Nieminen,
cond-mat/0304315 (2003)

\noindent
[18]
W. A. Harrison, Electronic Structure and the Properties of Solids: The Physics
of the Chemical Bond, W. H. Freeman and Company, San Francisco, 1980,
pp. 185-192. 

\noindent
[19]
J. K\"ubler, A. R. Williams, C. B. Sommers, Phys. Rev. B 28 (1983) 1745.

\end{document}